\documentclass[prl,amssymb,twocolumn,tightenlines,showpacs]{revtex4-1}

\usepackage{graphicx}
\usepackage{amsmath}
\usepackage{verbatim}
\usepackage{subfigure}

\begin{document}

\title{Realizing controllable noise in photonic quantum information channels}

\author{A. Shaham}
\affiliation{Racah Institute of Physics, Hebrew University of
Jerusalem, Jerusalem 91904, Israel}
\author{H.S. Eisenberg}
\affiliation{Racah Institute of Physics, Hebrew University of
Jerusalem, Jerusalem 91904, Israel}


\begin{abstract}
Controlling the depolarization of light is a long-standing open
problem. In recent years, many demonstrations have used the
polarization of single photons to encode quantum information. The
depolarization of these photons is equivalent to the decoherence
of the quantum information they encode. We present schemes for
building various depolarizing channels with controlled properties
using birefringent crystals. Three such schemes are demonstrated
and their effects on single photons are shown by quantum process
tomography to be in good agreement with a theoretical model.
\end{abstract}

\maketitle

Light depolarization is a fundamental optical phenomena. It was
studied as early as the nineteenth century, when measurement and
characterization methods of the polarization state of light were
introduced\cite{Stokes}. Methods for complete depolarization of
light, such as the Cornu and Lyot depolarizers, have been known
for many decades\cite{Hodgson,Lyot,Loeber_Lyot}. The Cornu and
wedge depolarizers require the light beam to be wide, as the loss
of coherence between the two polarizations is achieved via
averaging over the spatial degrees of freedom. In the case of the
Lyot depolarizer, short coherence length is required and the
averaging is over the temporal degrees of freedom. Nevertheless,
there are currently no methods that enable control of all of the
aspects of the depolarization process.

Quantum information is commonly encoded in the polarization of
single photons\cite{Nielsen}. Depolarization of such photons acts
as quantum noise on the stored information, i.e. the interaction
between the information encoding units and the environment results
in decoherence. In order to study quantum decoherence in general
and its effect on quantum information protocols in particular, it
is desirable to create quantum channels with controlled noise.
Such channels will be useful for testing quantum error correction
and quantum key distribution protocols\cite{BB84,Bruss}. Other
uses for these channels are to test for the existence of
decoherence-free subspaces\cite{Kwiat_Dephasing} and for
generating partially-mixed entangled
states\cite{Puentes_Entangled}.

In recent years, depolarizing channels were studied by several
methods. When a single birefringent crystal was used, only
dephasing channels were demonstrated, as we will show
later\cite{Kwiat_Dephasing}. Optical scatterers such as emulsions,
multi-mode fibers and ground glass, give variable depolarization,
but are also accompanied by a spread in $k$ space, resulting in
considerable loss when collected for further
processing\cite{Morgan,Puentes_Scatter,Puentes_Entangled}. Another
approach is to use polarization scramblers of various
kinds\cite{Billings,Karpinski}. These realizations are equivalent
to fast polarization rotations and averaging measurements over
times longer than the typical rotation periods. Nevertheless, when
used with single photons, each photon by itself is completely
polarized. Controllable depolarizers were demonstrated by using
two wedge depolarizers with variable beam diameter\cite{Nambu}, or
with a tunable relative angle\cite{Puentes_Wedge}. These channels
are hard to model and their anisotropy level is uncontrolled as
they couple the polarization with many spatial degrees of freedom.

In this letter we present a theoretical and experimental study of
various controllable depolarizing channels. We study channels that
are composed of a sequence of birefringent crystals and wave
plates. The depolarization and its anisotropy depends on the order
and relative angles between the channel components. The generated
channels are mostly anisotropic and can be tuned continuously
between no depolarization and complete dephasing. These channels
were characterized by the transmission of polarized single
photons, generated by spontaneous parametric down-conversion.
Quantum process tomography (QPT) was used to compare the
experimental results with theory\cite{Chuang,Kwiat_Tomo}.

The information in a classical channel can be degraded only by
bit-flip errors. Thus, such a channel is completely described by a
single parameter - the bit-flip error probability. In comparison,
quantum channels can have a constant unitary rotation and three
types of errors, represented by the Pauli operators: a bit-flip
that swaps the logical $|0\rangle$ and $|1\rangle$ amplitudes, a
phase flip between the amplitudes and the combination of the two,
which is a third orthogonal operation. Isotropic decoherence is
the case when the three error probabilities are equal.

A polarization qubit can be described either by a density matrix
operator $\hat{\rho}$ or equivalently, by a point in the
Poincar\'{e} sphere. The Cartesian coordinates of this point are
the Stokes parameters $\overline{S}=(S_1,S_2,S_3)$ which describe
the linear ($|h\rangle$,$|v\rangle$), diagonal
($|p\rangle=(|h\rangle+|v\rangle)/\sqrt{2}$,
$|m\rangle=(-|h\rangle+|v\rangle)/\sqrt{2}$) and circular
($|r\rangle=(|h\rangle+i|v\rangle)/\sqrt{2}$,
$|l\rangle=(i|h\rangle+|v\rangle)/\sqrt{2}$) polarization
components, respectively. The Degree Of Polarization (DOP) is
defined as D, the length of the Stokes vector\cite{Born}:
\begin{equation}\label{DOP}
D=\sqrt{S^2_1+S^2_2+S^2_3}\equiv\sqrt{1-4det(\hat{\rho})}.
\end{equation}
The perfectly polarized states are described by the surface of the
sphere ($D=1$) and its center designates the completely
unpolarized state ($D=0$). The inside of the sphere includes all
partially polarized states ($0<D<1$). The physical meaning of the
DOP is the ratio between the polarized light intensity and the
total light intensity.

Consider an arbitrarily polarized wave packet that is passing
through a birefringent crystal\cite{Loeber_Crystal}. The temporal
walk-off $\tau= L\frac {\Delta n}{c}$ between two wave packets,
each polarized along one of the symmetry axes of the crystal,
depends on the crystal length $L$, its refractive index difference
$\Delta n$ and the speed of light $c$. We assume that the
coherence time of the wave packets $t_c$ is shorter than the
walk-off $\tau$. If the light wave packet is not polarized
linearly along one of the crystal symmetry axes, its two
components acquire temporal distinguishability. Thus, the
polarization and temporal degrees of freedom become entangled. The
role of the environment in general decoherence models is fulfilled
here by the temporal degrees of freedom. As the detectors are
insensitive to the short temporal walk-offs, they cannot
distinguish between the wave packets, effectively tracing out the
temporal degrees of freedom. The result is an effective
depolarization since no coherence can be observed between the two
orthogonal polarizations. The depolarization operation is
described in the Poincar\'{e} sphere picture by a projection of
the initial Stokes vector on the direction that represents the
crystal principal axes. For example, a birefringent crystal
aligned along the $h$-$v$ directions will project any initial
state onto the $S_1$ direction. This kind of operation is referred
to as a 'dephasing channel'\cite{Nielsen}.

A single crystal configuration can apply any level of
depolarization to any initial linear polarization. On the other
hand, for such a configuration there is always another
polarization direction that experiences no depolarization
whatsoever. Thus, we consider a second crystal, that is placed
after the first one\cite{Loeber_Lyot}. For historical reasons, let
us first assume that the second crystal is twice as long as the
first one. The two crystals are coupling orthogonal temporal
degrees of freedom, as the first crystal couples  $t=0$ with
$\tau$, while the second couples $t=0$ with $2\tau$ and $t=\tau$
with $3\tau$. Thus, this configuration can be described as two
consecutive projections of the initial polarization state onto the
Stokes directions defined by the crystals' axes. In the case where
the orientation of the crystals differ by 45$^\circ$, the two
projections are perpendicular, resulting in a final state at the
sphere center ($D=0$, the completely unpolarized state) for any
initial state. This configuration in known as the 'Lyot
depolarizer'\cite{Lyot,Loeber_Lyot}.

The relevant error rates for practical tests of quantum
information protocols are below 20\%\cite{Kern}. For this reason,
it is desirable to have a depolarizing scheme that can be tuned to
small values of depolarization or even to zero depolarization.
Thus, we investigate a configuration of two identical crystals. If
the second crystal is oriented at $90^\circ$ with respect to the
first one, the polarization time delay created by the first
crystal is exactly compensated for by the second. For any other
relative angle, there can be up to three different temporal modes:
$t=0$ and $\tau$ are coupled by the first crystal, while the
second crystal couples between them and an additional third delay
$t=2\tau$. Changing the relative angle between the two crystals
affects the occupation of each of the three modes, which results
in a different depolarization. Hence, by tuning this angle we
control the channel depolarization level. The modal occupation can
be easily calculated for a given initial polarization and tuning
angle. The polarization two-dimensional density matrix
$\hat{\rho}_f$ of the final polarization state is the normalized
sum of the polarization states of all the occupied temporal modes
$|\psi_t\rangle$:
\begin{equation}\label{density_mat}
\ \hat{\rho}_f=\sum_{t=1}^T\alpha_t|\psi_t\rangle\langle\psi_t|\,,
\end{equation}
where $T$, $t$ and $\alpha_t$ are the number of relevant temporal
modes, their index and weights, respectively. This calculation is
equivalent to partially tracing over the temporal degrees of
freedom of the $2T$-dimensional polarization-time density matrix.
In addition, the effect of wave-plates is simple to calculate as
they rotate each of the temporal modes separately. With this
method, it is possible to calculate the final polarization state
for every initial state that passes through a sequence of
birefringent crystals (a "depolarizer"). From the resulting
density matrix it is possible to calculate the DOP by Eq.
\ref{DOP}.

\begin{figure}[tbp]
\includegraphics[angle=0,width=86mm]{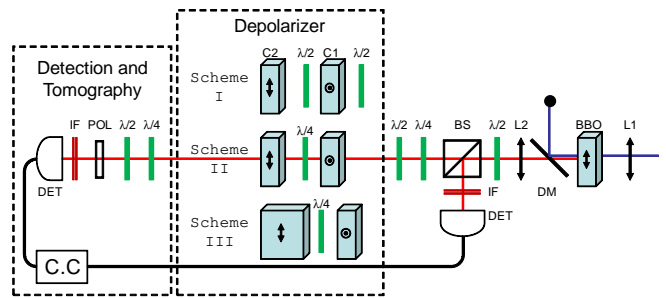}
\caption{\label{setup} (Color online) The experimental setup and
depolarizing schemes. See full description in text.}
\end{figure}

A 780\,nm Ti:Sapphire pulsed laser of 76\,MHz repetition rate was
frequency doubled and the 390\,nm pulses were focused into and
collinearly down-converted in a 1\,mm thick type-I BBO crystal
(See Fig. \ref{setup}). The down-converted signal was filtered by
a dichroic mirror (DM) and collimated with a lens (L2). One photon
of the pair was split by a beam-splitter (BS) and detected, and
the second photon was directed to the depolarizer. The
polarization state of the depolarized photons was characterized by
wave plates and a polarizer (POL). Photons were filtered by 5\,nm
bandpass filters (IF), corresponding to a coherence time of
$t_{c}\simeq180 \textrm{ fs}$, and then coupled into single-mode
fibers leading to single photon detectors (DET). We characterized
the depolarization of three initial states $|h\rangle$,
$|p\rangle$, and $|r\rangle$ which are mutually
unbiased\cite{Bruss}, by quantum state tomography
(QST)\cite{Kwiat_Tomo}. The $|v\rangle$ state was also measured as
required by QPT.

The first depolarizing scheme that we present is composed of two
2\,mm long Calcite crystals (C) with two $\lambda/2$ wave plates
(See Fig. \ref{setup}). The crystals are fixed perpendicularly,
with one wave plate before and the other after the first crystal.
Rotation of the wave plates in opposite directions by an angle of
$\theta/2$ is equivalent to the rotation of the first crystal by
$\theta$. When $\theta=0^\circ$ no depolarization occurs, and when
$\theta=90^\circ$ the depolarizer is equivalent to a dephasing
channel of a single crystal. Wave plates were used instead of
direct rotation of the crystal to eliminate an unwanted angle
dependant retardation.

\begin{figure}[tbp]
\includegraphics[angle=0,width=86mm]{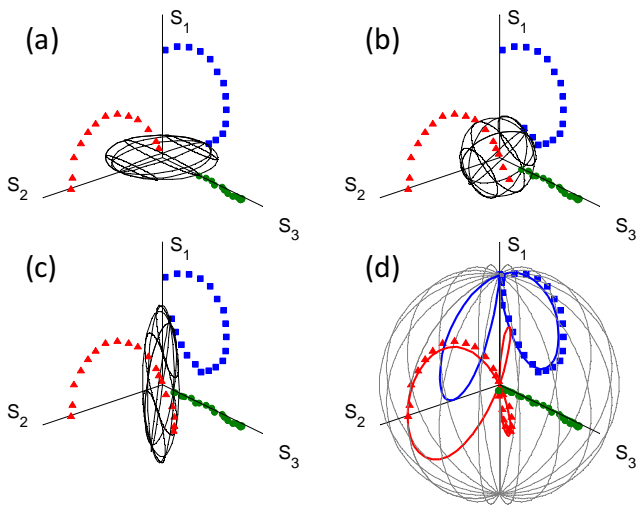}
\caption{\label{poincare} (Color online) Experimentally measured
final states for $|h\rangle$ (blue squares), $|p\rangle$ (red
triangles), and $|r\rangle$ (green circles) inputs, plotted for a
range of crystal angles up to a final value of (a)
$\theta=45^\circ$, (b) $\theta=54.74^\circ$, and (c)
$\theta=67.5^\circ$. The wire-mesh ellipsoids represent the
mapping of the surface of the Poincar\'{e} sphere by the
depolarizing operation as measured by QPT. (d) Comparison between
measurements in the range $0^\circ<\theta<90^\circ$ and the
theoretical model. Theoretical curves are presented as solid lines
in the range $0^\circ<\theta<180^\circ$.}
\end{figure}

\begin{figure}[tbp]
\includegraphics[angle=0,width=86mm]{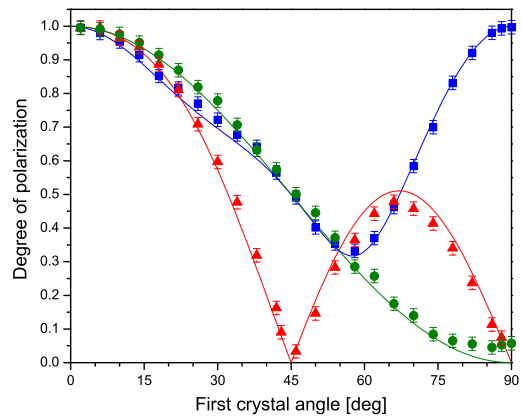}
\caption{\label{DOP1} (Color online) Experimentally measured
degree of polarization of the output states as a function of the
equivalent first crystal rotation angle $\theta$ for the first
depolarizing scheme. Initial states and their representation are
the same as in Fig. \ref{poincare}. Model predictions are
presented as solid lines.}
\end{figure}

The transformation of purely polarized states through the
depolarizer for various angles between the two crystals are
presented in Fig. \ref{poincare}. In Figs. \ref{poincare}(a-c),
the measured QPT mappings of the initial sphere are shown for
three specific cases. These cases are when the polarization of
only one initial state is completely lost ($\theta=45^\circ$),
when the channel is isotropic ($\theta=54.7^\circ$) and when two
initial states are depolarized identically ($\theta=67.5^\circ$).
The process fidelities, when compared to the model for all three
cases, were higher than 97\%. The measured output states for the
three mutually unbiased polarizations were obtained using the
maximal-likelihood QST algorithm\cite{Kwiat_Tomo}. They are also
presented, up to the respective angles. Figure \ref{poincare}(d)
compares these results with theory. The results show a very good
agreement with the model. From these results, we have calculated
the DOP as a function of $\theta$. The DOP results are presented
in Fig. \ref{DOP1} with their theoretical predictions. The three
special cases of Figs. \ref{poincare}(a-c) are clearly reproduced
in our measurements.

In general, this depolarizer is anisotropic, except for
$\theta=54.7^\circ=\tan^{-1}(\sqrt{2})$. Analytical calculation of
the DOP for this angle shows that the final value of 1/3 is
independent of the initial state, as can be seen in Fig.
\ref{DOP1}, where all three curves intersect. The isotropy of this
configuration is apparent in Fig. \ref{poincare}(b), where the
polarization sphere is mapped to another sphere of radius 1/3.

\begin{figure}[tbp]
\includegraphics[angle=0,width=86mm]{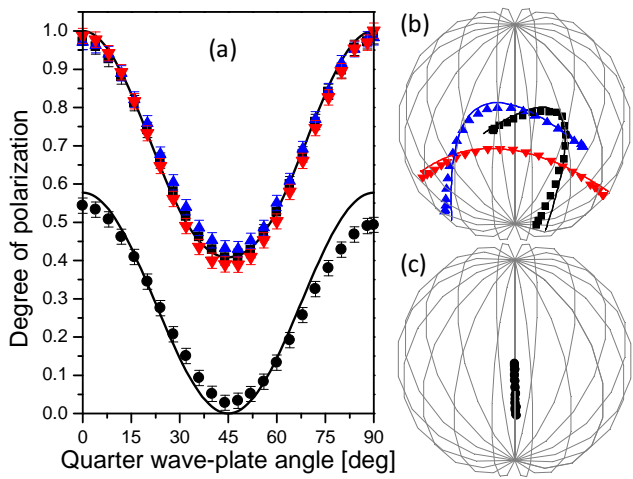}
\caption{\label{DOP2} (Color online) (a) Experimentally measured
degree of polarization as a function of the quarter-wave plate
angle $\theta$ for the second and the third depolarizing schemes.
For the second scheme, the three orthogonal initial states with
$S_1=-\sqrt{1/3}$ after depolarization are presented by black
squares, blue triangles and red inverted triangles. For the third
scheme, the depolarized states of all three initial states should
be identical. Results for one of these states are presented by
black circles. (b,c) The measured states in the Poincar\'{e}
sphere for the respective initial states.}
\end{figure}

We studied a second depolarizing scheme that was composed of two
perpendicularly fixed identical crystals with a quarter-wave plate
between them (See Fig. \ref{setup}). The quarter-wave plate angle
$\theta$ is set to be zero when the principal axes of the wave
plate and the first depolarizing crystal are parallel. for any
given initial polarization $(S_1,S_2,S_3)$, the final DOP is

\begin{eqnarray}\label{DOP_lambda_4}
D^2&=&\frac{1}{4}\left(\frac{19}{8}+\frac{3}{2}\cos\left(4\theta\right)+\frac{1}{8}\cos\left(8\theta\right)\right)\\
&+&\frac{S_{1}^{2}}{4}\left(-\frac{7}{8}+\frac{1}{2}\cos\left(4\theta\right)+\frac{3}{8}\cos\left(8\theta\right)\right).
\nonumber
\end{eqnarray}
All states with the same $|S_1|$ value, result in the same DOP for
a certain $\theta$. It is possible to find three mutually unbiased
polarization bases that have the same $S_1=\pm\frac{1}{\sqrt{3}}$
value, and thus, experience the same depolarization. We define
this situation as symmetric depolarization. For such bases, the
DOP can be tuned between 1 and $\frac{1}{\sqrt{6}}\approx0.41$ as
a function of $\theta$. We generated these states and
characterized them after passing through the second depolarizing
scheme. The DOP results are shown in Fig. \ref{DOP2}(a), and their
Poincar\'{e} representation in Fig. \ref{DOP2}(b). A good
agreement with theory is observed.

The third scheme adds the possibility for symmetric depolarization
down to complete depolarization. The difference between the second
and third schemes is the doubling of the second crystal's
thickness (See Fig. \ref{setup}). As before, the final DOP depends
only on the initial $S_1$ value and the wave plate angle $\theta$,
but now it takes values between 0 and $\frac{1}{\sqrt{3}}\approx
0.58$. This result is due to an effective additional $S_1$
projection to the output of the second scheme. At the
$\theta=45^\circ$ position, the third scheme is exactly a Lyot
depolarizer that completely depolarizes any initial polarization
by consecutive $S_1$ and $S_3$ projections.

Results for an initially polarized state with
$S_1=-\frac{1}{\sqrt{3}}$ are shown in Figs. \ref{DOP2}(a) and
\ref{DOP2}(c). As can be seen, when $\theta=45^\circ$, the state
is completely depolarized. The state tomography results (Fig.
\ref{DOP2}(c)) reveal the difference from the previous scheme as
an extra $S_1$ projection.

Although we have demonstrated the effects of the various
depolarizing schemes using single photons, these depolarizers
would also be effective on any classical light with short enough
coherence time. We repeated our measurements with laser pulses,
and demonstrated identical results (not presented here). Thus,
these results apply not only to polarization encoded qubits, but
to any classical scenario where controlled depolarization is
required.

In this work we use crystals that are long enough to completely
separate the two polarization components. It is possible to
deliberately use shorter crystals that will leave a portion of the
two wave packets overlapping. The same qualitative results will be
achieved this way, but with smaller magnitude. Another scheme that
enables control of the channel isotropy level is to combine three
setups of the same kind. The crystal lengths should be tripled for
the second setup and tripled again for the third, in order for
each of them to affect orthogonal temporal modes. If the second
setup is positioned at $45^\circ$ with respect to the first and a
quarter-wave plate at $45^\circ$ is placed between the second and
the third setups, then each setup will be oriented at a different
Stokes direction. This scheme can generate isotropic channels of
any depolarization amount.

In conclusion, we have demonstrated a scheme for the realization
of various quantum channels for photon polarization qubits with
controllable decoherence. Isotropic and anisotropic depolarization
processes are possible. Channels were characterized by QPT using
the maximal-likelihood algorithm. All the results are in a good
agreement with a simple theoretical model. These depolarizers can
be used to evaluate the performance of quantum error correction
and quantum key distribution protocols. In addition, they can be
utilized in any classical optics setup where controllable
depolarization is required. The authors would like to thank the
Israeli Science Foundation for supporting this work under grant
366/06.


\begin{thebibliography}{99}

\bibitem{Stokes}G.G. Stokes, Trans. Cambridge Phil. Soc. \textbf{9}, 399 (1852).

\bibitem{Hodgson}N. Hodgson and H. Weber, \emph{Laser Resonators and Beam Propagation: Fundamentals, Advanced Concepts and Applications} (Second edition), Springer (2005), p. 171-172.

\bibitem{Lyot}B. Lyot, Annales de l'Observatoire d'astronomie physique de Paris (Meudon) \textbf{8}, 102 (1929).

\bibitem{Loeber_Lyot}A.P. Loeber, J. Opt. Soc. Am. \textbf{72}, 650 (1982).

\bibitem{Nielsen}M.A. Nielsen and I.L. Chuang, \emph{Quantum Computation and Quantum Information} (Cambridge University Press, Cambridge, UK, 2000).

\bibitem{BB84}C.H. Bennett and G. Brassard, \textit{Proceedings of the IEEE International Conference on Computers, Systems and Signal Processing, Bangalore, India} (IEEE, New York, 1984), p. 175.

\bibitem{Bruss}D. Bru{\ss}, Phys. Rev. Lett. \textbf{81}, 3018 (1998).

\bibitem{Kwiat_Dephasing}P.G. Kwiat \textit{et al.},  Science \textbf{290}, 498 (2000).

\bibitem{Puentes_Entangled}G. Puentes \textit{et al.}, Phys. Rev. A \textbf{75}, 032319 (2007).

\bibitem{Morgan}S.P. Morgan, M.P. Khong and M.G. Somekh, Appl. Opt. \textbf{36}, 1560 (1997).

\bibitem{Puentes_Scatter}G. Puentes, D. Voigt, A. Aiello and J.P. Woerdman, Opt. Lett. \textbf{30}, 3216 (2005).

\bibitem{Billings}B.H. Billings,  J. Opt. Soc. Am. \textbf{41}, 966 (1951).

\bibitem{Karpinski}M. Karpi\'{n}ski, C. Radzewicz and K. Banaszek, J. Opt. Soc. Am. B \textbf{25}, 668 (2008).

\bibitem{Nambu}Y. Nambu \textit{et al.}, Proc. SPIE Int. Soc. Opt. Eng. \textbf{4917}, 13 (2002).

\bibitem{Puentes_Wedge}G. Puentes \textit{et al.}, Opt. Lett. \textbf{31}, 2057 (2006).

\bibitem{Chuang}I.L. Chuang and M.A. Nielsen, J. Mod. Opt. \textbf{44}, 2455 (1997).

\bibitem{Kwiat_Tomo}J.B. Altepeter, E.R. Jeffrey and P.G. Kwiat, Adv. At. Mol. Opt. Phys. \textbf{52}, 105 (2005).

\bibitem{Born}A. Al-Qasimi \textit{et al.}, Opt. Lett. \textbf{32}, 1015 (2007).

\bibitem{Loeber_Crystal}S. Lu and A.P. Loeber, J. Opt. Soc. Am. \textbf{65}, 248 (1975).

\bibitem{Kern}O. Kern and J.M. Renes, Quan. Inf. Comp. \textbf{8}, 756 (2008).
\end{thebibliography}
\end{document}